\documentclass[manuscript]{aastex}

\newcommand{\myemail}{ken.marsh@astro.cf.ac.uk}

\slugcomment{Accepted for publication in Astrophysical Journal}

\shorttitle{Parallaxes of Ultracool Brown Dwarfs}
\shortauthors{Marsh et al.}

\begin{document}

\title{Parallaxes and Proper Motions of Ultracool Brown Dwarfs \\
    of Spectral Types Y and Late T}

\author{Kenneth A. Marsh\altaffilmark{1,2}, Edward L. Wright\altaffilmark{3},
J. Davy Kirkpatrick\altaffilmark{1}, Christopher R. Gelino\altaffilmark{1},
Michael C. Cushing\altaffilmark{4}, 
Roger L. Griffith\altaffilmark{1}, Michael F. Skrutskie\altaffilmark{5},
Peter R. Eisenhardt\altaffilmark{6}}

\altaffiltext{1}{Infrared Processing and Analysis Center, California
Institute of Technology, Pasadena, CA 91125}
\altaffiltext{2}{School of Physics \& Astronomy, Cardiff University,
Cardiff CF24 3AA, UK; \myemail}
\altaffiltext{3}{Department of Physics \& Astronomy, UCLA, PO Box 951547,
Los Angeles, CA 90095}
\altaffiltext{4}{Department of Physcis \& Astronomy, The University of Toledo,
2801 West Bancroft Street, Toledo, OH 43606}
\altaffiltext{5}{Department of Astronomy, University of Virginia, 
Charlottesville, VA 22904}
\altaffiltext{6}{NASA Jet Propulsion Laboratory, 4800 Oak Grove Drive,
Pasadena, CA 91109}

\begin{abstract}
We present astrometric measurements of eleven nearby ultracool brown dwarfs of
spectral types Y and late-T, based on imaging observations from a variety
of space-based and ground-based telescopes. These measurements have been used
to estimate relative parallaxes and proper motions via maximum likelihood 
fitting of geometric model curves.  To compensate for the modest
statistical significance ($\stackrel{<}{_\sim} 7$) of our parallax
measurements we have employed a novel Bayesian procedure for distance estimation
which makes use of an {\em a priori\/} distribution of tangential
velocities, $V_{\rm tan}$, assumed similar to that implied by previous
observations of T dwarfs.  Our estimated distances are therefore somewhat 
dependent on that assumption.  Nevertheless, 
the results have yielded distances for five of our eight Y dwarfs and
all three T dwarfs.  Estimated distances in all cases are 
$\stackrel{>}{_\sim}3$ pc.  
In addition, we have obtained significant estimates
of $V_{\rm tan}$ for two of the Y dwarfs; both are $<100$ km s$^{-1}$,
consistent with membership in the thin disk population.
Comparison of absolute magnitudes with model
predictions as a function of color shows that the Y dwarfs are significantly 
redder in $J\!-\!H$ than predicted by a cloud-free model.
\end{abstract}


\section{Introduction}

Determining accurate distances to brown dwarfs is important for a number of 
reasons. Firstly, distance is a vital quantity in establishing not only the 
space density of these objects, but also the luminosity function which
can then be used to test models of star formation at the lowest masses. 
Secondly, distances allow the spectra of brown dwarfs to be placed on an 
absolute flux scale to provide more quantitative checks of atmospheric models. 
Thirdly, 
distances for the nearest objects allow us to construct a more complete view of
our own Solar Neighborhood, allowing us to directly visualize the relative 
importance of brown dwarfs in the Galactic context. Sometimes, distance 
determinations produce results wholly unanticipated. For example, the $J$-band 
overluminosity of the T4.5 dwarf 2MASS J05591914$-$1404488 (Figure 2 of
Dahn et al.\ 2002) was unexpected despite its location near the 
$J$-band bump at the L/T transition (e.g., 
Looper et al.\ 2008), a feature thought to be associated with decreasing
cloudiness \citep{mar2010}. It has been suggested, however, that the 
overluminosity is due to the presence of
an unresolved binary \citep{burg2002,dup2012}.  Similarly unexpected 
was the recent determination that young, field L dwarfs are often significantly 
underluminous for their spectral types at near-infrared magnitudes 
\citep{fahe2012}.

Some of the earliest parallax determinations for brown dwarfs were by  
\citet{dahn2002}, \citet{tinn2003}, \citet{vrba2004}, once surveys such as the 
Two Micron All-Sky Survey (2MASS; Skrutskie et al.\ 2006), the Sloan Digital 
Sky Survey (SDSS; York et al.\ 2000),
and the Deep Near-infrared Survey of the southern sky 
(DENIS; Epchtein et al.\ 1997) began to identify L and T dwarfs in large 
numbers. More recently, parallax programs by groups such as \citet{maro2010} 
and \citet{dup2012} have pushed astrometry measurements to the 
latest T spectral subclasses. With the discovery of Y dwarfs from WISE 
(Cushing et al.\ 2011; Kirkpatrick et al.\ 2012) we are now pushing these 
measurements to even colder temperatures \citep{chas2012}. In this 
paper we present distance and/or proper motion measurements for an 
additional eight Y dwarfs, along with three nearby late-T dwarfs, and
present the first tangential velocity measurements for Y dwarfs.

\section{Observations}

Our set of objects includes all known Y dwarfs for which we have imaging data 
at a sufficient number of epochs for parallax and proper motion estimation.
The exception is WISE 1828+2650, presented separately by
\citet{chas2012}.  In addition, we have included three late T dwarfs from
an investigation of the low-mass end of the substellar mass function
within 8 pc of the sun \citep{davy2011}.  The complete sample is listed
in the observing log shown in Table~\ref{tbl-1}.  

Each of these objects has been observed at two or three epochs by the
Wide-field Infrared Survey Explorer (WISE; \citet{wright2010}) and at least 
four more epochs
of imaging observations by the IRAC instrument \citep{faz2004} on the {\em Spitzer 
Space Telescope\/} ({\em Spitzer\/}; \citet{wer2004}), the WFC3 instrument \citep{stra2011} of the 
{\em Hubble Space
Telescope\/} (HST), and various ground-based observatories. The observatories 
and instruments used are listed in the footnote of Table~\ref{tbl-1}, and 
further details are given by \citet{davy2011,davy2012}. 

\section{Astrometry Measurement Procedure}

Astrometric information was extracted from the observed images at the various
epochs using the standard maximum likelihood technique in which a point spread
function (PSF) is fit to each observed source profile.  The technique was 
essentially the same as used in 2MASS, the details of
which are given by \citet{cutri2003}, except that the source
extraction results presented here were made using coadded images rather than
focal-plane images.  The positional uncertainties
were estimated using an error model which includes the effects of instrumental
and sky background noise and PSF uncertainty.  The PSF and its associated
uncertainty map were estimated for each image individually using a set of
bright stars in the field, the median number for which was 14.
Since the coadded images were Nyquist
sampled or better, sinc interpolation was appropriate during PSF estimation
and subsequent profile fitting to the data.

In order to minimize the systematic effects of
focal-plane distortion and plate scale and rotation errors, our
astrometry is based on relative positions using a reference star (or set
of reference stars) in the vicinity of the object.  
For most objects we were able to find a reference star within $\sim10''$ 
common to all images except for those of WISE, due to the lower sensitivity of 
the latter.  In order to incorporate the WISE data it has
therefore been necessary to include 
bright reference stars which in general were much more widely separated
from the brown dwarf (up to $\sim100''$).  Most of these were taken from the 2MASS
Point Source Catalog \citep{cutri2003}.  
In order not to let these stars significantly compromise
the astrometry measurements from the more sensitive images with close
reference stars, we used a hybrid scheme in which the bright stars were
treated as secondary references,  
bootstrapped to the close reference stars using the images in which they
were in common.

The procedure is based on the following measurement model for the observed
separation between the brown dwarf and reference star:

\begin{eqnarray}
\alpha_t - \alpha_{it}^{\rm ref} & = & \alpha_t^{\rm BD} - (\alpha_i^{\rm cat}
+ \Delta\alpha_i^{\rm cat}) + \nu_t - \nu_{it} \label{eq1} \\
\delta_t - \delta_{it}^{\rm ref} & = & \delta_t^{\rm BD} - (\delta_i^{\rm cat}
+ \Delta\delta_i^{\rm cat}) + \nu'_t - \nu'_{it} 
\label{eq2}
\end{eqnarray}
where $\alpha_t,\delta_t$ and 
$\alpha_{it}^{\rm ref},\delta_{it}^{\rm ref}$ 
represent the extracted positions of the brown dwarf and $i$th reference star, 
respectively, 
estimated from the image at epoch $t$ based on the nominal position calibration
of that image;  $\alpha_i^{\rm cat},\delta_i^{\rm cat}$ represent the
catalog position of the reference star, and 
$\Delta\alpha_i^{\rm cat},\Delta\delta_i^{\rm cat}$ represent errors in the
catalog position;
$\nu_{t},\nu'_{t}$ represent the estimation errors for the brown dwarf, and
$\nu_{it},\nu'_{it}$ represent the estimation errors for the reference star.
These estimation errors include the effects of random measurement noise 
on the source extraction as
well as the residual effects of focal-plane distortion in the position
differences. We assume that they can all be described by 
zero-mean Gaussian random processes.

If we further assume that the $\Delta\alpha_i^{\rm cat}, 
\Delta\delta_i^{\rm cat}$ 
are described {\em a priori\/} by zero-mean Gaussian random processes
with standard deviations substantially larger than the extraction uncertainties
of the reference stars, then an optimal estimate of the brown dwarf position can
be obtained from:

\begin{eqnarray}
  \hat{\alpha_t}^{\rm BD} & = & \alpha_t + \frac{1}{N_t}
   \sum_{i\in{\cal R}(t)} (\alpha_i^{\rm cat} - \alpha_{it}^{\rm ref}) \\
  \hat{\delta_t}^{\rm BD} & = & \delta_t + \frac{1}{N_t}
   \sum_{i\in{\cal R}(t)} (\delta_i^{\rm cat} - \delta_{it}^{\rm ref})
\end{eqnarray}
where ${\cal R}(t)$ is the set of detected reference stars in the image at 
epoch $t$, and $N_t$ is the number of stars in the set.

The resulting estimates are included in Table \ref{tbl-1} in the form
of offsets from the nominal position of the brown dwarf at each epoch, and the
set of reference stars used is given in Table \ref{tbl-2}.
After having obtained $\hat{\alpha_t}^{\rm BD}$
and $\hat{\delta_t}^{\rm BD}$, the individual reference star
catalog errors can then be estimated using:

\begin{eqnarray}
  \hat{\Delta\alpha_i}^{\rm cat} & = & -\alpha_i^{\rm cat} +
      \frac{1}{N_i}\sum_{t\in {\cal E}(i)} (\hat{\alpha_t}^{\rm BD}  + 
    \alpha_{it}^{\rm ref} - \alpha_t)  \\ 
  \hat{\Delta\delta_i}^{\rm cat} & = & -\delta_i^{\rm cat} +
      \frac{1}{N_i}\sum_{t\in {\cal E}(i)} \,(\,\hat{\delta_t}^{\rm BD}  + 
    \delta_{it}^{\rm ref} - \,\delta_t)
\end{eqnarray}
where ${\cal E}(i)$ is the set of all epochs for which the $i$th reference star 
is detected in the corresponding image, and $N_i$ is the number of epochs in 
the set.

These values can be applied as corrections to the catalog positions of the
reference stars, enabling a corresponding time series of estimated brown dwarf positions
to be obtained separately for each individual reference
star via (\ref{eq1}) and (\ref{eq2}).  The scatter in 
these estimates
then provides a check on the assumptions regarding systematic effects
such as focal-plane distortion and possible small proper motions of the
reference stars themselves.  We have included the effect of this scatter
in the final quoted error bars in Table \ref{tbl-1}.  

\section{Estimation of Parallax and Proper Motion}

The measurement model incorporated the effects of parallax and linear
proper motion, with approprate correction for the Earth-trailing orbit in
the case of {\em Spitzer\/} observations.  The equations used 
\citep{davy2011} were as follows:

\begin{eqnarray}
\cos\delta_1 (\alpha_i-\alpha_1) & = & \Delta\alpha + \mu_\alpha (t_i-t_1) + \pi_{\rm trig} \vec{R}_i \cdot \hat{W} \\
\delta_i-\delta_1 & = & \Delta\delta +\mu_\delta (t_i-t_1) - \pi_{\rm trig} \vec{R}_i \cdot \hat{N}
\end{eqnarray}

\noindent where $t_i$ is the observation time [yr] of the $i$th astrometric 
measurement, and $R_i$ is the vector
position of the observer relative to the Sun in celestial
coordinates and astronomical units. $\hat{N}$ and $\hat{W}$
are unit vectors pointing North and West from the position of the source.
$R_i$ is the position of the Earth for 2MASS, SDSS, WISE, and HST observations;
for {\em Spitzer\/} observations, $R_i$ is the position of the spacecraft.
The observed positional difference on the left hand side is in arcsec,
the parameters $\Delta\alpha$ and $\Delta\delta$ are in arcsec,
the proper motion $\mu_\alpha$ and $\mu_\delta$ are in arcsec/yr,
and the parallax $\pi_{\rm trig}$ is in arcsec.

\begin{figure}
\epsscale{1.0}
\plotone{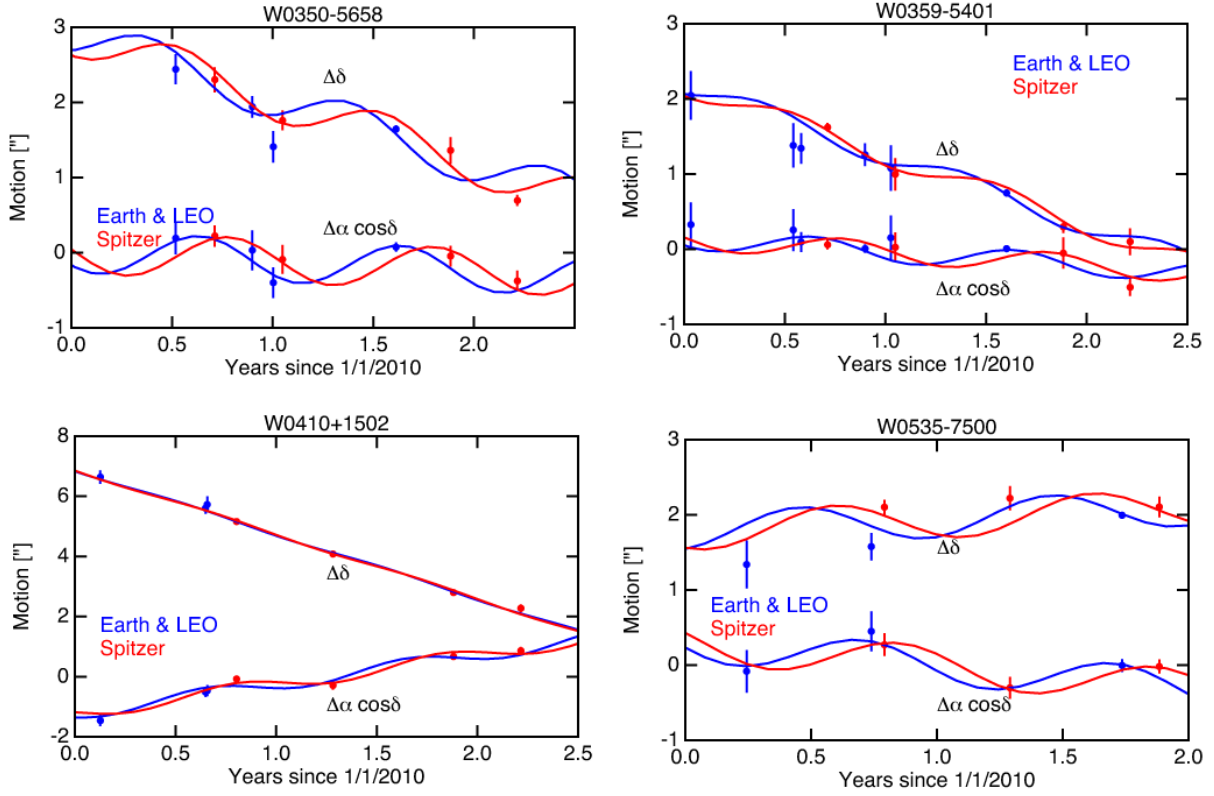}
\caption{Proper motion and parallax fits to astrometry measurements of four
of the Y dwarfs. Blue symbols represent observations from the
ground and Low Earth Orbit (LEO), and red symbols represent
{\em Spitzer\/} observations. The blue and red curves
represent the corresponding model fits, respectively.
The origins for the position offsets on the vertical (Motion) axes have been 
adjusted with respect to the values in Table \ref{tbl-1};  the $\Delta\delta$
and $\Delta\alpha\cos\delta$ values are relative to a constant position
fit, so they are relative to the weighted mean of the $\alpha$ and $\delta$.
In addition, the $\Delta\delta$ values are offset for clarity by different
amounts for the different plots.
\label{fig1a}}
\end{figure}

\begin{figure}
\epsscale{1.0}
\plotone{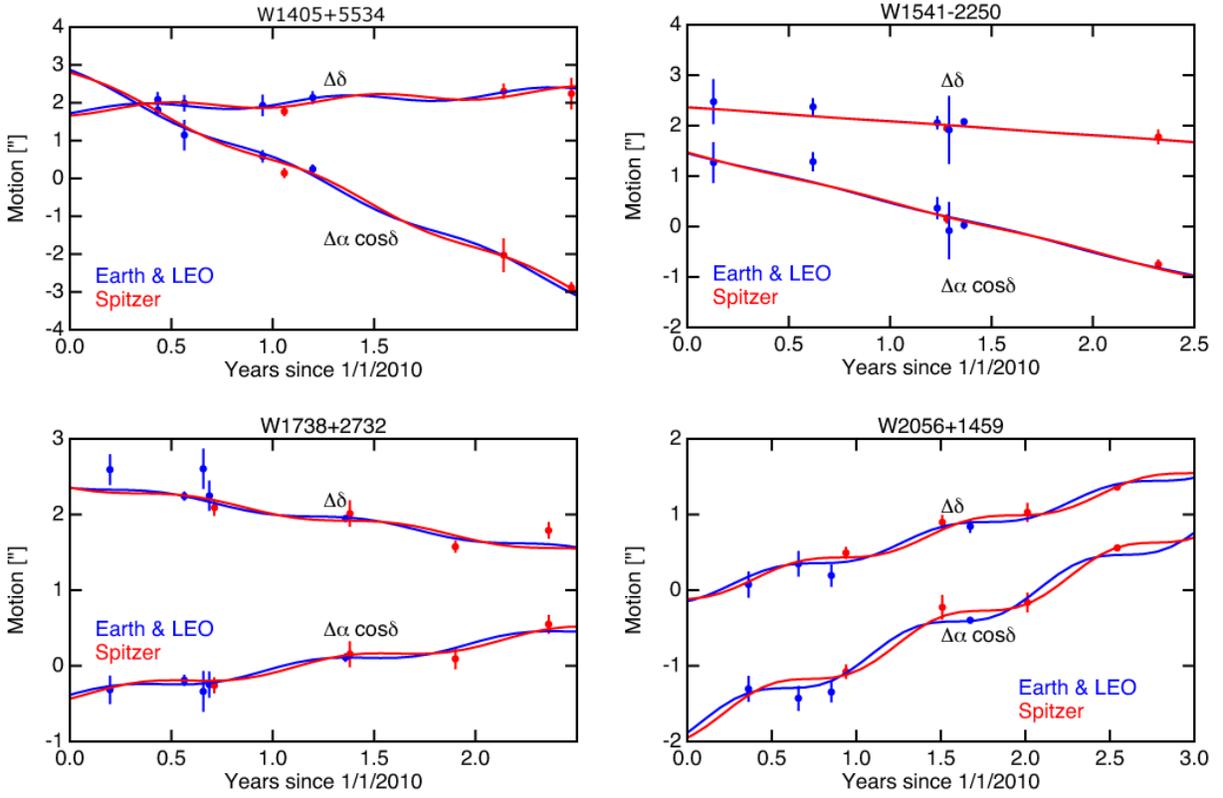}
\caption{Proper motion and parallax fits to astrometry measurements of the
remaining four Y dwarfs.  Color convention is the same as for
Figure \ref{fig1a}.
\label{fig1b}}
\end{figure}

\begin{figure}
\epsscale{0.6}
\plotone{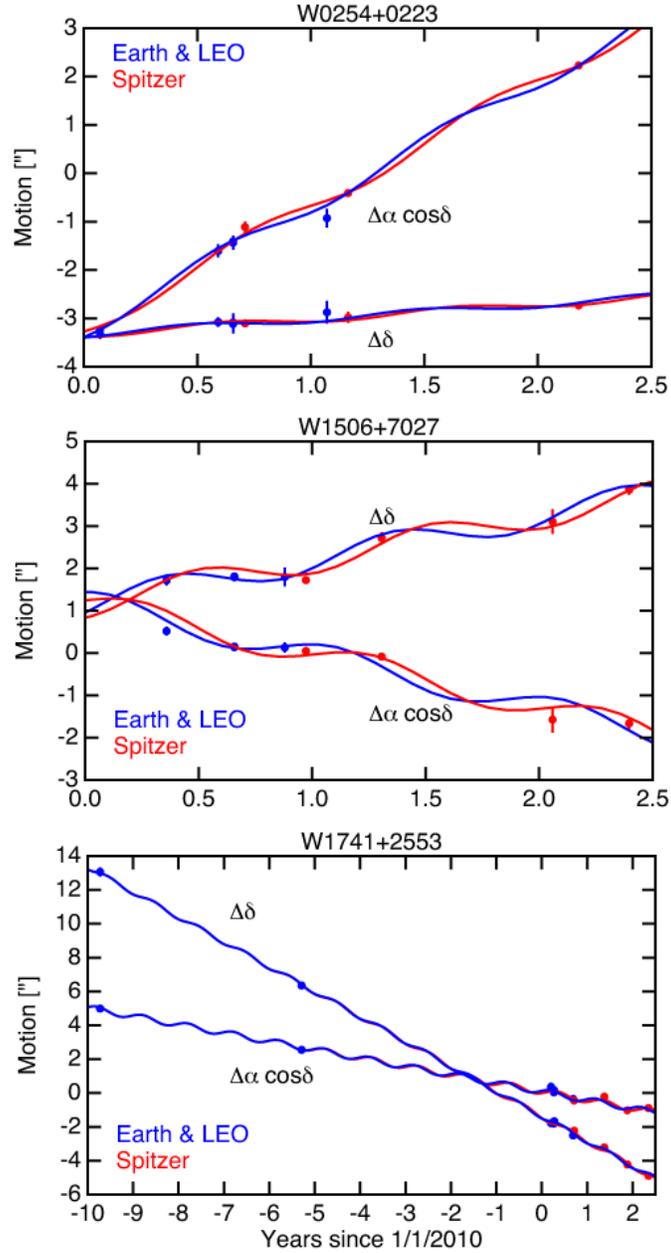}
\caption{Proper motion and parallax fits to astrometry measurements of the 
three T dwarfs. Color convention is the same as for Figure 
\ref{fig1a}.
\label{fig1c}}
\end{figure}

Maximum likelihood estimates, based on the assumption
of Gaussian measurement noise, were made of five parameters:  the RA and Dec
position offsets of the source at a specified reference time, the RA and Dec
rates of proper motion, and the parallax. 
The uncertainties were derived using the standard procedure for 
maximum likelihood estimation \citep{wha71} using the positional uncertainties
quoted in Table \ref{tbl-3}.
The resulting estimates of proper motion and parallax and their associated
uncertainties are given in 
Table \ref{tbl-3}, and the model fits with respect to the astrometry
measurements are presented in Figures \ref{fig1a}, \ref{fig1b}, and
\ref{fig1c}.  
The chi squared values, $\chi^2$, for the parameter fits in Table \ref{tbl-3} 
are, for the most part, close to the number of degrees of freedom, 
$N_{\rm df}$, indicating
reasonably good modeling of position uncertainties.  Formally, the 
probability of exceeding $\chi^2$ given $N_{\rm df}$ has a median value
0.29.

The parallaxes that we present are, strictly speaking, {\rm relative\/}
parallaxes since no correction has been made for the small parallaxes
and proper motions of the reference stars, most of which are relatively nearby.
However, the expected correction for such effects is only $\sim2$ mas
\citep{dup2012} which is at least an order of magnitude smaller than our 
typical astrometric uncertainties listed in Table \ref{tbl-3}, so in this
error regime the distinction between relative and absolute parallaxes is
unimportant.

In order to check to what extent our parallax and proper motion estimates
may have been affected by systematic effects of focal-plane distortion not
properly modeled by the statistical assumptions of the previous section, we 
have compared the rms residuals of the above fits (obtained using multiple 
reference stars) with those obtained using a single reference star
for each brown dwarf, and found that there was no significant difference.
This suggests that whatever residual focal plane distortion errors exist,
they are smaller than the random errors of source extraction.

We have converted our maximum likelihood estimates of parallax into most 
probable estimates of distance taking into account both the parallax 
measurements themselves and prior information.  The latter includes an 
assumption that our objects are spatially distributed in a statistically 
uniform manner.  Formally, that would imply that parallax values are 
distributed {\em a priori\/} as $P(\pi)\propto \pi^{-4}$; the singularity 
at zero would then lead to
difficulties in estimating the {\em a posteriori\/} most probable $\pi$.
Even though the zero parallax can be excluded on
physical grounds, there is still a bias towards small values such
that for $S/N < 4$, maximum likelihood parallax estimates become
insignificant \citep{lk73}. Fortunately there is additional prior 
information to alleviate this problem; small parallaxes 
(i.e. large distances) can be excluded if they are inconsistent with the 
observed proper motion based on an assumed velocity dispersion
of the objects being studied \citep{thor03}.

With these considerations in mind, our estimates of distance, $d$, are based 
on the following assumptions:
\smallskip

1. Our maximum likelihood parallax values, $\pi_{\rm ML}$, are distributed as 
Gaussians with standard deviation $\sigma_\pi$.

2. Our objects are distributed spatially in a statistically uniform way, so
that the {\em a priori\/} probability density distribution of $d$ 
is proportional to $d^2$.

3. The distribution of tangential velocities of Y dwarfs in the solar 
neighborhood can be described by a Gaussian random process with mean 
and standard deviation
$\bar{V}$ and $\sigma_V$ respectively; 
we assume the values $\bar{V}=30$ km s$^{-1}$ and
$\sigma_V=20$ km s$^{-1}$ respectively, representative of previous 
observations of T dwarfs \citep{faherty09}.
\smallskip

We then obtain the most probable distance, $\hat{d}$, by maximizing the 
conditional probability density $P(d|\pi_{\rm ML},\mu_{\rm ML})$, which
from Bayes' rule can be expressed by:
\begin{equation}
P(d|\pi_{\rm ML},\mu_{\rm ML}) \propto d^2 \exp(-(\mu_{\rm ML}d - \bar{V})^2/(2\sigma_V^2))
\exp(-(\pi_{\rm ML} - 1/d)^2/(2\sigma_\pi^2)) 
\label{eq9}
\end{equation}
where $\mu_{\rm ML}$ represents the magnitude of our maximum likelihood
estimate of proper motion.
Our distance estimates are presented in column 9 of Table \ref{tbl-3}.
The error bars correspond to the 0.159 and 0.841 points of the cumulative
distribution with respect to $P(d|\pi_{\rm ML},\mu_{\rm ML})$.

\section{Discussion}

As is evident from Table \ref{tbl-1}, our observations represent a mixed bag
in terms of telescopes (and hence spatial resolution) and time sampling
since they
were not specifically designed for astrometry, but rather
for followup photometry of brown dwarfs detected by WISE.
The quality of the observations was quite varied, and not always
with sufficient pixel subsampling for the estimation of the high
quality PSFs necessary for astrometry.  In the case of {\em Spitzer\/},
for example, each observation consisted of a set of only five dithered images.

In addition, the time sampling of the parallactic cadence is
of key importance in the estimation of parallax.  The ideal sampling
involves observations at solar elongation angles of $\pm90^\circ$, and
this is achieved by WISE, albeit with large position errors (typically
$\sim0.1-0.3''$). These elongation angles are critical for an object on the
ecliptic and less important at high ecliptic latitudes.  
For the parallax measurements described here,
the worst example of poor sampling was
WISE 1541-2250 for which all of the non-WISE observations were in one
quadrant of solar elongation angle (see column 8 of Table \ref{tbl-1}), 
so it is not surprising that the
observations did not yield a significant parallax measurement.  The
previous measurement, corresponding to an estimated distance range of 
2.2--4.1 pc \citep{davy2011}, was based on even fewer observations and
furthermore used position estimates for which the PSF errors were somewhat
underestimated.  Our present result of $>6$ pc therefore supercedes that 
estimate, but this object should clearly be revisited once a more 
optimal sampling of the parallactic ellipse has been obtained. By and
large, however, there is a good correlation between the sampling cadence and 
the quality of the parallax estimate;
future observations will be optimized both for image quality and
cadence.

Nevertheless, significant parallaxes ($S/N>3$) have been obtained for 
five of the eight Y dwarfs and all three of the T dwarfs, thus providing 
distance estimates.  Also, we have combined the latter with our proper
motion estimates to yield
tangential velocities, $V_{\rm tan}$.  Of course,
our estimated values,  $\hat{d}$ and $\hat{V}_{\rm tan}$, are somewhat 
dependent on the assumed prior distribution of $V_{\rm tan}$ in 
Eq. (\ref{eq9}), and the assumed similarity to the T dwarf
distribution may not be valid if the Y dwarfs represent a significantly
older population.  In order
to assess the sensitivity to this assumption, the distance estimates
were repeated using a $\sigma_V$ of 100 km s$^{-1}$.  It was found
that for a parallax significance $S/N>4$, the increase in $\sigma_V$
led to no more than a 20\% change (always in the positive direction)
in $\hat{d}$ and hence $\hat{V}_{\rm tan}$.  For lower values of $S/N$,
$\hat{V}_{\rm tan}$ becomes biased towards the {\em a priori\/}
value, $\bar{V}$, in Eq. (\ref{eq9}).  Thus in Table \ref{tbl-3} we 
quote $\hat{V}_{\rm tan}$ 
values only for $S/N>4$. Similarly, for $S/N<4$ the reliability of our distance 
estimates is dependent on the validity of the assumptions regarding the 
{\em a priori\/} distribution of $V_{\rm tan}$.

On this basis we obtained significant values of $V_{\rm tan}$ for two
of our Y dwarfs; 
both are $<100$ km s$^{-1}$, suggesting membership in
the thin disk population \citep{dup2012}.
Similar analysis techniques, both in terms of the
source extraction and parallax estimation, were used by \citet{wright2012}
to estimate the distance to the T8.5 object WISE 1118+3125, inferred (with
the aid of its observed common proper motion) to be a 
member of the $\xi$ UMa system, with excellent 
agreement with the known distance of that system.

\begin{figure}
\epsscale{0.8}
\plotone{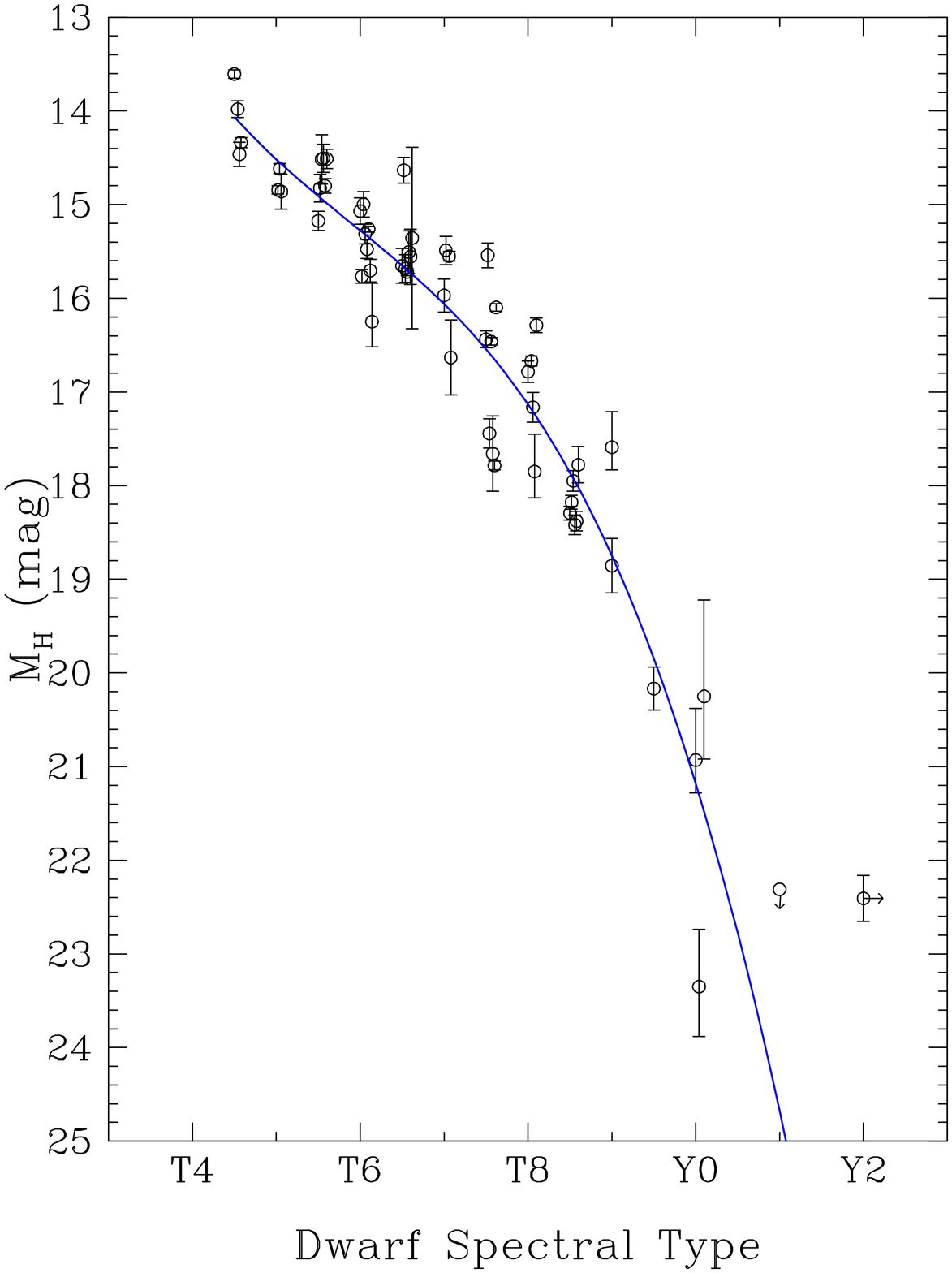}
\caption{Absolute H magnitude as a function of spectral type.  This is
a revised version of the corresponding figure in
\citet{davy2012} and includes the objects from the present
paper and the new parallax estimate for WISE 1828+2650 \citep{chas2012}.  
The blue curve represents the relation used by \citet{davy2012}, which appears
still to be an accurate representation of the absolute magnitude versus
spectral type trend despite the fact that our results have been revised since
the Kirkpatrick et al. paper was published.
\label{fig2}}
\end{figure}

\begin{figure}
\epsscale{0.8}
\plotone{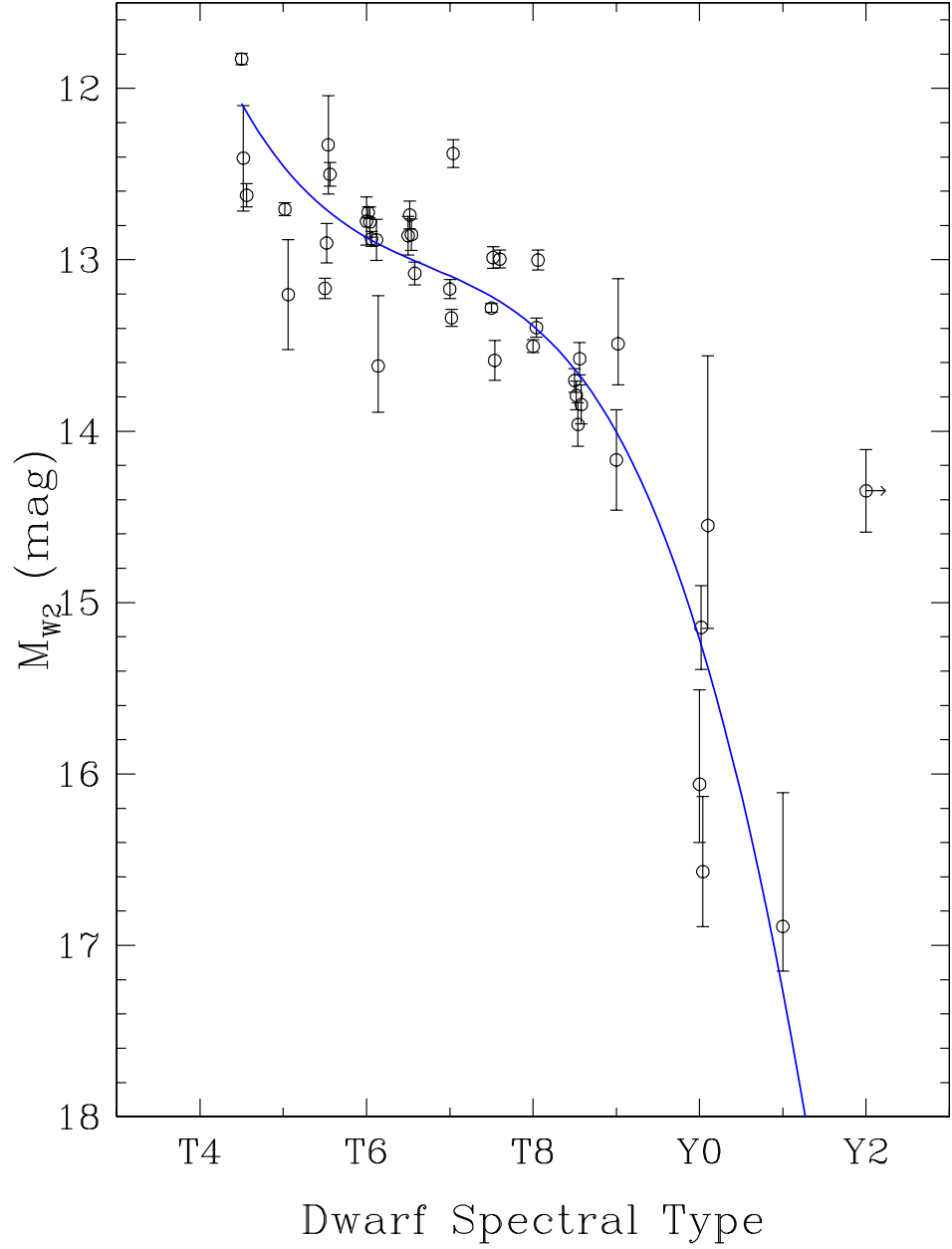}
\caption{Absolute W2 magnitude as a function of spectral type. As with
Figure \ref{fig2} it is taken from \citet{davy2012} except for the
inclusion of the objects from the present paper.  It also includes
WISE 1639-6847 (Tinney et al. 2012, submitted).
\label{fig3}}
\end{figure}

The distance estimates for the present sample,
all of which are $\stackrel{>}{_\sim}3$ pc, have enabled the estimation of 
absolute magnitudes.  These indicate that luminosities plummet at 
the T/Y boundary \citep{davy2012} as illustrated by
Figures \ref{fig2} and \ref{fig3} which represent updated versions of the
absolute magnitude versus spectral type plots from the latter work.
The steep decrease may at least partially account for the apparent scatter 
in absolute magnitudes of objects of the same spectral type, since in the 
Y0 regime an error of half a spectral type apparently corresponds to more than 
a magnitude difference in luminosity.  More data will be required to make
a definitive statement, however.

\begin{figure}
\epsscale{0.5}
\plotone{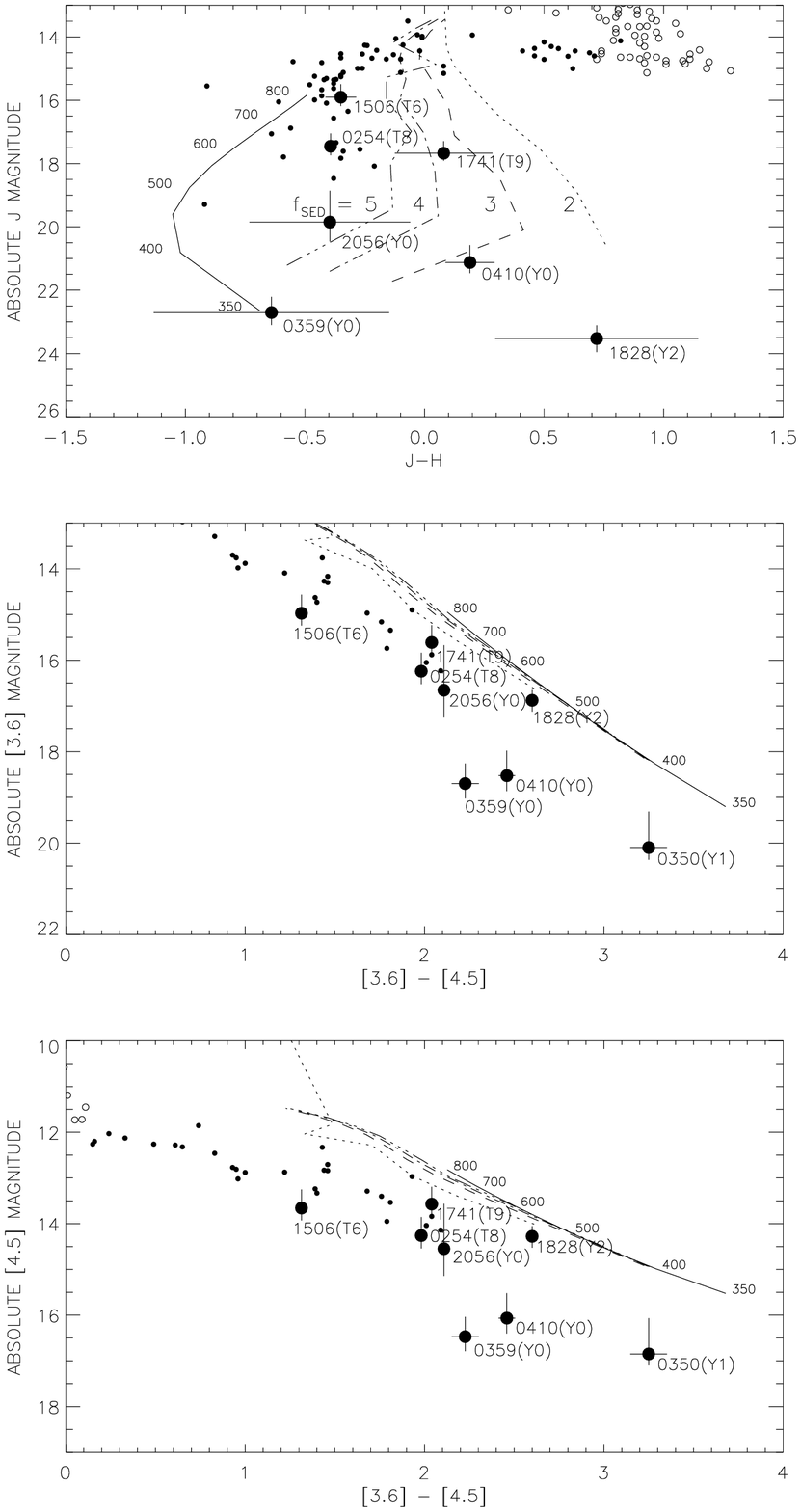}
\caption{Absolute magnitude as a function of color.  Large filled circles with
error bars represent the objects from this paper, plus WISE 1828+2650 
\citep{chas2012}. Also included are the L and T dwarfs from \citet{dup2012},
represented by open circles and small filled circles, respectively.  For 
comparison, model curves are overplotted.  The solid curve represents
a cloud-free model from \citet{s&m2008}, assuming $g=1000$ m s$^{-2}$, 
$K_{zz}=0$.  The numbers along this line represent the assumed values of 
effective temperature [K].  Also plotted (dashed/dotted lines) are four cloudy 
models from \citet{morley2012} with the same assumed $g$ and $K_{zz}$, and
with various values of the sedimentation
efficiency parameter, $f_{\rm sed}$, as indicated.
\label{fig4}}
\end{figure}

The absolute magnitudes also provide valuable guidance for models in the
ultra-cool regime.  To this end we have compared our observational results
with model-based and empirical predictions using plots of absolute 
magnitude as a function
of color, as shown in Figure \ref{fig4}.  The $M_J$ verus $J\!-\!H$ plot
in the upper panel shows that the Y dwarfs
continue the trend set by the L and T dwarfs based on the 
parallax observations of \citet{dup2012}. A key feature
is the turnover in the blueward progression of the color at $M_J\sim16$, 
at considerably redder $J\!-\!H$ than predicted by cloud-free
models \citep{s&m2008} as illustrated by the solid curve.
Such behavior is also apparent in the color-magnitude diagrams for cloud-free
models presented by \citet{leg2010}.  The dotted/dashed curves in Figure
\ref{fig4} represent models incorporating the effect of clouds containing
various amounts of Cr, MnS, Na$_2$S, ZnS, and KCl condensates
\citep{morley2012}, as indicated by the sedimentation efficiency parameter,
$f_{\rm sed}$; lower values correspond to optically thicker clouds.
It is apparent that these models can account at least partly for the 
relative redness of some of the $J\!-\!H$ colors but they predict a blueward 
hook for temperatures below 400 K which does not appear to be matched by the 
observations. Perhaps some of the scatter in $J\!-\!H$ colors in
Figure \ref{fig4} might be explained in terms of a patchy cloud model;
it is also possible that the inclusion of water clouds might 
improve consistency with the observations.

Figure \ref{fig4} does show reasonable consistency between observations
and models based on IRAC colors, i.e., $M_{[3.6]}$ and $M_{[4.5]}$
as a function of the [3.6]-[4.5] color.  The only major discrepancy is
that WISE 1828+2650, whose effective temperature is believed to be
$\sim300$ K, falls at a location more indicative of 500 K on these plots.

\acknowledgments

We thank C. Morley for providing the results of model calculations, 
and also the referee for very helpful comments.
This publication makes use of data products from the Wide-field Infrared
Survey Explorer, which is a joint project of the University of California,
Los Angeles, and the Jet Propulsion Laboratory / California Institute
of Technology, funded by the National Aeronautics and Space Administration.
This work is based in part on observations made
with the Spitzer Space Telescope, which is operated by the Jet
Propulsion Laboratory, California Institute of Technology, under a
contract with NASA.  Support for this work was provided by NASA through
an award issued to programs 70062 and 80109 by JPL/Caltech.  This work is also
based in part on observations made with the NASA/ESA Hubble Space
Telescope, obtained at the Space Telescope Science Institute, which is
operated by the Association of Universities for Research in Astronomy, Inc.,
under NASA contract NAS 5-26555.  These observations are associated with
program 12330, support for which was provided by NASA through a grant from the
Space Telescope Science Institute.
This paper also includes data gathered with the 6.5 meter Magellan Telescopes
located at Las Campanas Observatory, Chile.
This research has made use of the NASA/IPAC Infrared Science Archive
(IRSA), which is operated by the Jet Propulsion Laboratory, California
Institute of Technology, under contract with the National Aeronautics
and Space Administration, and also the SIMBAD database, operated at CDS,
Strasbourg, France.

\clearpage

\begin{deluxetable}{lccclccrcc}
\tabletypesize{\scriptsize}
\tablecolumns{10}
\tablewidth{0pc}
\rotate
\tablecaption{Observing log and relative astrometry measurements}
\tablehead{
\colhead{Object} & \colhead{Sp} & \colhead{RA (nom)} & \colhead{Dec (nom)} & \colhead{Instr.}   & \colhead{Band}    & \colhead{Date} & \colhead{Elong.} &
\colhead{$\Delta\alpha\cos\delta$} & \colhead{$\Delta\delta$}\\
  & &[$^\circ$] &[$^\circ$] & & & & [$^\circ$]\ \ \ \  & [$''$] & [$''$] }
\startdata
 WISE J035000.32-565830.2 & Y1 &     57.501375 &    -56.975006 & WISE   & W2 & 2010-07-09 &  -89.9 &   -0.153 (0.232) &   -0.062 (0.208) \\
 & & & & Spitzer & [4.5] & 2010-09-18 & -158.2 &   -0.131 (0.119) &   -0.126 (0.156) \\
 & & & & PANIC  & J & 2010-11-25 &  134.2 &   -0.314 (0.279) &   -0.562 (0.182) \\
 & & & & WISE   & W2 & 2011-01-02 &   95.5 &   -0.743 (0.221) &   -1.094 (0.215) \\
 & & & & Spitzer & [4.5] & 2011-01-19 &   78.2 &   -0.927 (0.309) &   -1.486 (0.273) \\
 & & & & HST    & J & 2011-08-13 & -123.2 &   -0.271 (0.094) &   -0.857 (0.062) \\
 & & & & Spitzer & [4.5] & 2011-11-20 &  139.5 &   -0.378 (0.129) &   -1.148 (0.131) \\
 & & & & Spitzer & [4.5] & 2012-03-20 &   17.0 &   -0.722 (0.137) &   -1.808 (0.075) \\
 WISE J035934.06-540154.6 & Y0 &     59.892083 &    -54.031703 & WISE   & W2 & 2010-01-13 &   93.4 &   -0.203 (0.298) &   -0.200 (0.316) \\
 & & & & WISE   & W2 & 2010-07-18 &  -89.2 &   -0.273 (0.278) &   -0.867 (0.287) \\
 & & & & PANIC  & J & 2010-08-01 & -102.6 &   -0.434 (0.145) &   -0.907 (0.166) \\
 & & & & Spitzer & [4.5] & 2010-09-18 & -148.9 &   -0.468 (0.267) &   -0.632 (0.251) \\
 & & & & PANIC  & H & 2010-11-25 &  143.5 &   -0.521 (0.052) &   -0.987 (0.144) \\
 & & & & WISE   & W2 & 2011-01-11 &   95.7 &   -0.374 (0.295) &   -1.169 (0.301) \\
 & & & & Spitzer & [4.5] & 2011-01-19 &   87.5 &   -0.500 (0.196) &   -1.284 (0.237) \\
 & & & & HST    & J & 2011-08-09 & -110.0 &   -0.521 (0.039) &   -1.499 (0.040) \\
 & & & & Spitzer & [4.5] & 2011-11-20 &  148.8 &   -0.564 (0.194) &   -1.969 (0.085) \\
 & & & & Spitzer & [4.5] & 2012-03-20 &   26.3 &   -1.036 (0.113) &   -2.147 (0.178) \\
 WISEP J041022.71+150248.5 & Y0 &     62.594667 &     15.046819 & WISE   & W2 & 2010-02-16 &   96.3 &   -0.001 (0.188) &   -0.052 (0.222) \\
 & & & & WISE   & W2 & 2010-08-26 &  -89.2 &    0.945 (0.168) &   -1.083 (0.193) \\
 & & & & WIRC   & J & 2010-08-29 &  -92.1 &    0.997 (0.172) &   -0.965 (0.258) \\
 & & & & Spitzer & [4.5] & 2010-10-21 & -144.0 &    1.411 (0.066) &   -1.538 (0.050) \\
 & & & & Spitzer & [4.5] & 2011-04-14 &   39.7 &    1.175 (0.134) &   -2.621 (0.090) \\
 & & & & Spitzer & [4.5] & 2011-11-19 & -172.8 &    2.134 (0.134) &   -3.894 (0.079) \\
 & & & & Spitzer & [4.5] & 2012-03-20 &   63.7 &    2.320 (0.126) &   -4.417 (0.132) \\
 WISE J053516.80-750024.9 & $\geq$Y1 &     83.820042 &    -75.007019 & WISE   & W2 & 2010-03-31 &  -89.5 &   -0.361 (0.284) &    0.458 (0.317) \\
 & & & & WISE   & W2 & 2010-09-28 &   95.9 &    0.172 (0.266) &    0.693 (0.182) \\
 & & & & Spitzer & [4.5] & 2010-10-17 &   77.1 &   -0.006 (0.151) &    1.221 (0.099) \\
 & & & & Spitzer & [4.5] & 2011-04-17 & -106.0 &   -0.582 (0.145) &    1.337 (0.160) \\
 & & & & HST    & J & 2011-09-27 &   97.1 &   -0.284 (0.086) &    1.112 (0.036) \\
 & & & & Spitzer & [4.5] & 2011-11-20 &   43.3 &   -0.297 (0.093) &    1.223 (0.137) \\
 WISEPC J140518.40+553421.5 & Y0p?  &    211.326667 &     55.572628 & WISE   & W2 & 2010-06-08 &   96.0 &   -0.165 (0.145) &    0.107 (0.198) \\
 & & & & WIRC   & J & 2010-07-26 &   50.2 &   -0.828 (0.401) &    0.013 (0.216) \\
 & & & & WISE   & W2 & 2010-12-14 &  -88.8 &   -1.388 (0.161) &   -0.050 (0.284) \\
 & & & & Spitzer & [4.5] & 2011-01-22 & -128.5 &   -1.829 (0.130) &   -0.210 (0.125) \\
 & & & & HST    & J & 2011-03-14 & -180.0 &   -1.723 (0.118) &    0.155 (0.171) \\
 & & & & Spitzer & [4.5] & 2012-02-21 & -158.6 &   -4.002 (0.445) &    0.323 (0.203) \\
 & & & & Spitzer & [4.5] & 2012-06-22 &   82.1 &   -4.862 (0.148) &    0.260 (0.412) \\
 WISE J154151.65-225024.9 & Y0.5 &    235.465250 &    -22.840358 & WISE   & W2 & 2010-02-17 &  -89.8 &    0.206 (0.532) &   -0.209 (0.708) \\
 & & & & WISE   & W2 & 2010-08-15 &   96.4 &    0.041 (0.175) &   -0.093 (0.173) \\
 & & & & FIRE   & J & 2011-03-27 & -127.5 &   -0.959 (0.196) &   -0.351 (0.198) \\
 & & & & Spitzer & [4.5] & 2011-04-13 & -144.3 &   -1.019 (0.120) &   -0.418 (0.140) \\
 & & & & NEWFIRM & J & 2011-04-17 & -148.2 &   -1.340 (0.566) &   -0.430 (0.682) \\
 & & & & MMIRS  & J & 2011-05-14 & -174.4 &   -1.260 (0.113) &   -0.292 (0.128) \\
 & & & & Spitzer & [4.5] & 2012-04-28 & -159.7 &   -2.028 (0.101) &   -0.568 (0.144) \\
 WISE J173835.53+273259.0 & Y0 &    264.648083 &     27.549758 & WISE   & W2 & 2010-03-14 &  -90.8 &   -0.136 (0.188) &    0.036 (0.204) \\
 & & & & WIRC   & J & 2010-07-26 &  139.5 &   -0.009 (0.074) &   -0.314 (0.060) \\
 & & & & WIRC   & J & 2010-08-29 &  106.9 &   -0.154 (0.271) &    0.046 (0.266) \\
 & & & & WISE   & W2 & 2010-09-09 &   96.3 &   -0.064 (0.174) &   -0.309 (0.200) \\
 & & & & Spitzer & [4.5] & 2010-09-18 &   87.5 &   -0.075 (0.103) &   -0.465 (0.110) \\
 & & & & HST    & J & 2011-05-12 & -148.5 &    0.291 (0.056) &   -0.613 (0.048) \\
 & & & & Spitzer & [4.5] & 2011-05-20 & -156.2 &    0.336 (0.171) &   -0.544 (0.174) \\
 & & & & Spitzer & [4.5] & 2011-11-26 &   19.1 &    0.276 (0.138) &   -0.982 (0.082) \\
 & & & & Spitzer & [4.5] & 2012-05-12 & -149.2 &    0.734 (0.124) &   -0.765 (0.109) \\
 WISEPC J205628.90+145953.3 & Y0 &    314.120417 &     14.998147 & WISE   & W2 & 2010-05-13 &  -90.6 &    0.027 (0.172) &    0.042 (0.167) \\
 & & & & WIRC   & J & 2010-08-29 &  166.0 &   -0.100 (0.168) &    0.316 (0.170) \\
 & & & & WISE   & W2 & 2010-11-08 &   96.1 &   -0.015 (0.135) &    0.163 (0.144) \\
 & & & & Spitzer & [4.5] & 2010-12-10 &   63.8 &    0.354 (0.148) &    0.437 (0.160) \\
 & & & & Spitzer & [4.5] & 2011-07-06 & -142.0 &    1.087 (0.229) &    0.853 (0.146) \\
 & & & & HST    & J & 2011-09-04 &  160.5 &    0.933 (0.035) &    0.810 (0.083) \\
 & & & & Spitzer & [4.5] & 2012-01-06 &   36.5 &    1.152 (0.203) &    0.936 (0.231) \\
 & & & & Spitzer & [4.5] & 2012-07-18 & -154.2 &    1.889 (0.032) &    1.332 (0.042) \\
\hline
 WISEPA J025409.45+022359.1 & T8 &     43.539375 &      2.399750 & WISE   & W2 & 2010-01-27 &   94.9 &    0.052 (0.085) &   -0.745 (0.119) \\
 & & & & WISE   & W2 & 2010-08-05 &  -90.7 &    1.673 (0.139) &   -0.509 (0.100) \\
 & & & & WIRC   & H & 2010-08-29 & -113.7 &    1.839 (0.144) &   -0.514 (0.181) \\
 & & & & WIRC   & J & 2010-08-29 & -113.7 &    1.864 (0.120) &   -0.546 (0.189) \\
 & & & & Spitzer & [4.5] & 2010-09-17 & -132.2 &    2.164 (0.110) &   -0.536 (0.064) \\
 & & & & WISE   & W2 & 2011-01-27 &   95.1 &    2.349 (0.193) &   -0.306 (0.231) \\
 & & & & Spitzer & [4.5] & 2011-03-02 &   60.8 &    2.868 (0.068) &   -0.411 (0.126) \\
 & & & & Spitzer & [4.5] & 2012-03-07 &   55.0 &    5.504 (0.049) &   -0.162 (0.067) \\
 WISEPC J150649.97+702736.0 & T6 &    226.708208 &     70.460000 & WISE   & W2 & 2010-05-12 &   95.4 &   -0.101 (0.106) &   -0.010 (0.118) \\
 & & & & WIRC   & H & 2010-08-29 &   -9.0 &   -0.479 (0.083) &    0.064 (0.065) \\
 & & & & WIRC   & J & 2010-08-29 &   -9.0 &   -0.467 (0.082) &    0.086 (0.059) \\
 & & & & WISE   & W2 & 2010-11-18 &  -89.0 &   -0.488 (0.119) &    0.069 (0.220) \\
 & & & & Spitzer & [4.5] & 2010-12-22 & -123.5 &   -0.573 (0.069) &   -0.005 (0.065) \\
 & & & & Spitzer & [4.5] & 2011-04-23 &  114.0 &   -0.707 (0.077) &    0.990 (0.136) \\
 & & & & Spitzer & [4.5] & 2012-01-23 & -155.8 &   -2.192 (0.307) &    1.377 (0.292) \\
 & & & & Spitzer & [4.5] & 2012-05-25 &   82.4 &   -2.278 (0.148) &    2.130 (0.115) \\
 WISEPA J174124.26+255319.5 & T9 &    265.351083 &     25.888750 & 2MASS  & J & 2000-04-11 & -117.8 &   -0.069 (0.138) &    0.188 (0.104) \\
 & & & & SDSS   & z & 2004-09-16 &   90.1 &    2.320 (0.087) &    8.248 (0.104) \\
 & & & & WISE   & W2 & 2010-03-15 &  -90.7 &    0.114 (0.261) &    0.125 (0.099) \\
 & & & & PAIRITEL & H & 2010-04-09 & -115.4 &   -0.194 (0.082) &    0.109 (0.109) \\
 & & & & FanMt  & J & 2010-04-10 & -116.4 &   -0.136 (0.038) &    0.231 (0.058) \\
 & & & & FanMt  & H & 2010-04-10 & -116.4 &   -0.103 (0.100) &    0.240 (0.064) \\
 & & & & WISE   & W2 & 2010-09-10 &   96.4 &   -0.582 (0.132) &   -0.596 (0.184) \\
 & & & & Spitzer & [4.5] & 2010-09-18 &   88.6 &   -0.679 (0.143) &   -0.334 (0.179) \\
 & & & & Spitzer & [4.5] & 2011-05-20 & -155.1 &   -0.463 (0.184) &   -1.317 (0.200) \\
 & & & & Spitzer & [4.5] & 2011-11-20 &   26.3 &   -1.265 (0.052) &   -2.312 (0.059) \\
 & & & & Spitzer & [4.5] & 2012-05-08 & -144.2 &   -1.132 (0.088) &   -2.977 (0.073) \\
\enddata
\tablecomments{ The columns represent the object name, spectral type
from \citet{mikec2011,davy2012}, nominal 
RA and Dec position (J2000), the instrument (or telescope), band, UT date of observation, solar elongation angle, and
the measured positional offsets (in RA and Dec) of the source from its nominal 
position. The key to the entries in the Instrument column is as follows: \\
WISE = Wide-field Infrared Survey Explorer \citep{wright2010};  \\
HST = WFC3 camera on the Hubble Space Telescope \citep{stra2011}; \\
Spitzer = Infrared Array Camera (IRAC) on {\em Spitzer\/} \citep{wer2004}; \\
FanMt = Fan Mountain Near-infrared Camera (FanCam) \citep{kann2009}; \\
FIRE = Folded-port Infrared Echellette at Las Campanas Observatory 
\citep{sim2008,sim2010}; \\
MMIRS = MMT and Magellan Infrared Spectrograph \citep{mcleo2004}; \\
NEWFIRM = NOAO Extremely Wide-Field Infrared Imager at Cerro Tololo
\citep{swat2009}; \\
PANIC = Persson's Auxiliary Nasmyth Infrared Camera at Las Campanas Observatory
 \citep{martini2004}; \\
PAIRITEL = Peters Automated Infrared Imaging Telescope on Mt. Hopkins
\citep{bloom2006}; \\
SDSS = Sloan Digital Sky Survey \citep{york2000} \\
2MASS = 2 Micron All Sky Survey \citep{cutri2003} \\
WIRC = Wide-field Infrared Camera on the 5-m Hale Telescope \citep{wil2003}. \\
 }
\label{tbl-1}
\end{deluxetable}

\begin{deluxetable}{cccccc}
\tabletypesize{\scriptsize}
\tablecolumns{6}
\tablewidth{0pc}
\tablecaption{Reference stars used}
\tablehead{
\colhead{Object} & \colhead{Sp} & \colhead{RA(ref)} & \colhead{Dec(ref)} & 
\colhead{Separation}    & \colhead{Comment} \\
  & &  [$^\circ$] & [$^\circ$] & [$''$] & } 
\startdata
WISE 0350-5658 & Y1 &  57.505458 & -56.976833 &   10.4 &      \\
           &    &  57.498042 & -56.985000 &   36.6 &2MASS \\
           &    &  57.493500 & -56.961917 &   49.6 &2MASS \\
           &    &  57.469708 & -56.975861 &   62.2 &2MASS \\
           &    &  57.520292 & -56.993056 &   74.8 &2MASS \\
           &    &  57.539625 & -56.972556 &   75.6 &2MASS \\
WISE 0359-5401 & Y0 &  59.895458 & -54.033444 &    9.5 &      \\
           &    &  59.894458 & -54.021056 &   38.7 &2MASS \\
           &    &  59.908333 & -54.042639 &   52.3 &2MASS \\
           &    &  59.932417 & -54.040583 &   91.1 &2MASS \\
WISE 0410+1502 & Y0 &  62.600125 &  15.058056 &   44.7 &2MASS \\
           &    &  62.580167 &  15.039056 &   57.6 &2MASS \\
           &    &  62.607333 &  15.034722 &   62.0 &2MASS \\
           &    &  62.618333 &  15.049167 &   82.7 &2MASS \\
           &    &  62.607083 &  15.023306 &   95.0 &2MASS \\
           &    &  62.622125 &  15.043389 &   96.3 &2MASS \\
WISE 0535-7500 & $\geq$Y1 &  83.824208 & -75.009278 &    9.0 &      \\
           &    &  83.793542 & -75.004417 &   26.4 &2MASS \\
           &    &  83.811542 & -74.998583 &   31.4 &2MASS \\
           &    &  83.771250 & -75.010028 &   46.7 &2MASS \\
           &    &  83.769292 & -75.004389 &   48.2 &2MASS \\
           &    &  83.823500 & -74.990444 &   59.7 &2MASS \\
WISE 1405+5534 & Y0p? & 211.327083 &  55.574778 &    7.8 &      \\
           &    & 211.343208 &  55.584722 &   55.0 &      \\
           &    & 211.305583 &  55.585333 &   62.7 &      \\
           &    & 211.273042 &  55.574167 &  109.3 &2MASS \\
           &    & 211.380417 &  55.576639 &  110.4 &2MASS \\
WISE 1541-2250 & Y0.5 & 235.464417 & -22.836833 &   13.0 &2MASS \\
           &    & 235.466750 & -22.831694 &   31.6 &2MASS \\
           &    & 235.473958 & -22.848833 &   42.0 &2MASS \\
           &    & 235.477375 & -22.845306 &   44.0 &2MASS \\
           &    & 235.482167 & -22.843861 &   57.5 &2MASS \\
           &    & 235.467875 & -22.857306 &   61.6 &2MASS \\
WISE 1738+2732 & Y0 & 264.643542 &  27.547750 &   16.2 &2MASS \\
           &    & 264.657750 &  27.554444 &   35.2 &2MASS \\
           &    & 264.640292 &  27.535667 &   56.5 &2MASS \\
           &    & 264.657750 &  27.534028 &   64.5 &2MASS \\
           &    & 264.640917 &  27.530556 &   72.8 &2MASS \\
           &    & 264.652417 &  27.572083 &   81.5 &2MASS \\
WISE 2056+1459 & Y0 & 314.117042 &  15.000111 &   13.7 &2MASS \\
           &    & 314.118667 &  15.002556 &   17.0 &2MASS \\
           &    & 314.120417 &  15.007694 &   34.3 &2MASS \\
           &    & 314.132917 &  14.993361 &   46.8 &2MASS \\
           &    & 314.106625 &  14.999250 &   48.1 &2MASS \\
           &    & 314.123833 &  15.014750 &   60.9 &2MASS \\
\hline
WISE 0254+0223 & T8 &  43.540792 &   2.412833 &   47.3 &2MASS \\
           &    &  43.537250 &   2.386722 &   47.5 &2MASS \\
           &    &  43.557958 &   2.400333 &   66.9 &2MASS \\
           &    &  43.512667 &   2.395778 &   97.1 &2MASS \\
WISE 1506+7027 & T6 & 226.736375 &  70.461806 &   34.5 &2MASS \\
           &    & 226.677125 &  70.475250 &   66.4 &2MASS \\
           &    & 226.750958 &  70.443639 &   78.3 &2MASS \\
           &    & 226.658042 &  70.478833 &   90.8 &2MASS \\
WISE 1741+2553 & T9 & 265.355375 &  25.896583 &   31.4 &2MASS \\
           &    & 265.341375 &  25.893556 &   35.9 &2MASS \\
           &    & 265.332083 &  25.883611 &   64.2 &2MASS \\
           &    & 265.360417 &  25.905361 &   67.1 &2MASS \\
           &    & 265.346958 &  25.869194 &   71.7 &2MASS \\
           &    & 265.339750 &  25.905861 &   71.7 &2MASS \\
\enddata
\tablecomments{Columns represent the object name, spectral type,
the RA and Dec values of the
associated reference stars, their separations from the object, and a
comment column indicating which of the reference stars are in the Two Micron
All-Sky Survey (2MASS) point source catalog.}
\label{tbl-2}
\end{deluxetable}

\begin{deluxetable}{cccccccccc}
\tabletypesize{\scriptsize}
\tablecolumns{10}
\tablewidth{0pc}
\tablecaption{Parallax and Proper Motion Estimates}
\tablehead{
\colhead{Object} & \colhead{Sp} & \colhead{$\chi^2$} & \colhead{$N_{\rm df}$} &
\colhead{$\mu_\alpha\cos\delta$}    & \colhead{$\mu_\delta$} &
\colhead{$\pi$} & \colhead{Signif.}   & \colhead{d} &  $V_{\rm tan}$ \\
& & & & [$\,''$ yr$^{-1}$] & [$\,''$ yr$^{-1}$] & [$''$] & [sigmas] & [pc] &  [km s$^{-1}$] }
\startdata
WISE 0350-5658 &  Y1  & 14.22 & 11 & -0.125$\pm$0.097 & -0.865$\pm$0.076 & 0.291$\pm$0.050 & 5.8 & $3.7^{+1.6}_{-0.4}$  & $18\pm4$ \\
WISE 0359-5405 & Y0   & 13.02 & 15 & -0.177$\pm$0.053 & -0.930$\pm$0.062 & 0.145$\pm$0.039 & 3.7 & $5.9^{+1.3}_{-0.8}$   &  \\
WISE 0410+1502 & Y0   & 11.53  & 9  & 0.974$\pm$0.079  & -2.144$\pm$0.072 & 0.233$\pm$0.056 & 4.2 & $4.2^{+1.2}_{-0.6}$  & $50\pm10$\\
WISE 0535-7500 & $\geq$Y1   & 11.80 & 7  & -0.310$\pm$0.128 & 0.159$\pm$0.092  & 0.250$\pm$0.079 & 3.2 & $21^{+13}_{-11}$   &  \\
WISE 1405+5534 & Y0p?   & 9.16  & 9  & -2.297$\pm$0.096 & 0.212$\pm$0.137  & 0.133$\pm$0.081 & 1.6 & $>3.4$  & \\
WISE 1541-2250 & Y0.5   & 15.21 & 9  & -0.983$\pm$0.111 & -0.276$\pm$0.116 & -0.021$\pm$0.094 & $<1$ & $>6.0$  & \\
WISE 1738+2732 & Y0   & 15.22 & 13 &  0.348$\pm$0.071 & -0.354$\pm$0.055 & 0.066$\pm$0.050 & 1.3 & $>6.0$  & \\
WISE 2056+1459 & Y0   & 6.64 & 11 &  0.881$\pm$0.057 & 0.544$\pm$0.042  & 0.144$\pm$0.044 & 3.3 & $7.5^{+4.3}_{-1.8} $   &  \\
\hline
WISE 0254+0223 & T8   & 5.67 & 11 &  2.578$\pm$0.042 & 0.309$\pm$0.050  & 0.185$\pm$0.042 & 4.4 & $4.9^{+1.0}_{-0.6}$  & $62\pm10$ \\
WISE 1506+7027 & T6   & 17.44 & 11 & -1.241$\pm$0.085 & 1.046$\pm$0.064  & 0.310$\pm$0.042 & 7.4 & $3.4^{+0.7}_{-0.4}$  & $27\pm4$ \\
WISE 1741+2553 & T9   & 9.90 & 19 & -0.495$\pm$0.011 & -1.472$\pm$0.013 & 0.176$\pm$0.026 & 6.8 & $5.8^{+1.1}_{-0.6}$  & $45\pm6$ \\
\enddata
\tablecomments{Columns represent the object name, spectral type, chi squared
of the parallax/proper motion fit to the estimated positions, number of
degrees of freedom, proper motion in RA and Dec, the maximum likelihood estimate
of parallax and its statistical significance, most probable distance (corrected for Lutz-Kelker bias),
and the tangential velocity.  Distance lower limits are based on a $2\sigma$ criterion.  Tangential velocities are quoted only for cases with parallax
significance $>4$, otherwise the $V_{\rm tan}$ estimate 
becomes strongly biased towards the assumed {\em a priori\/} mean value of 30 
km s$^{-1}$.}
\label{tbl-3}
\end{deluxetable}

\end{document}